\documentclass[prl,twocolumn,aps,showpacs,preprintnumbers]{revtex4}
\usepackage{graphicx}

\usepackage{dcolumn}
\usepackage{bm}

\begin{document}
\newcommand{\Par}[1]{\left(#1\right)}
\newcommand{\om}{\omega}
\newcommand{\e}{\varepsilon}

\title{Suppression of Visibility in a Two-Electron Mach-Zehnder
Interferometer}

\author{Ya.~M.~Blanter$^{1,2}$ and Yuval Gefen$^2$}
 \affiliation{$^1$Kavli Institute of Nanoscience, Delft University of
 Technology, Lorentzweg 1, 2628 CJ Delft, The Netherlands\\
 $^2$ Department of Condensed Matter Physics, The Weizmann Institute
 of Science, Rehovot 76100, Israel}

\date{\today}

\begin{abstract}
We investigate the suppression of the visibility of Aharonov-Bohm
oscillations in a two-electron Mach-Zehnder interferometer that
leaves the single-electron current unchanged. In the case when the
sources emit either spin-polarized or entangled electrons, partial
distinguishability of electrons (coming from two different sources) suppresses
the visibility. 
Two-particle entanglement may produce behavior similar to
"dephasing" of two-particle interferometry.
\end{abstract}

 \maketitle

Aharonov-Bohm (AB) interferometry is a centerpiece in studies of
nanoscopic electronic systems. Of special interest are Mach-Zehnder
interferometers (MZIs)~\cite{Heiblum}: The absence of backscattering gives
rise to a large interference signal (large visibility), and allows
measurements in strong magnetic fields.

One conceptual step beyond single-particle interferometry is
``two-particle interferometry'', tailored after the Hanbury Brown
-- Twiss experiment. In particular, one can utilize a two-electron
MZI which features two current sources and two detectors (Fig.~1).
Such a device, suitable for the measurement of current
cross-correlations in the detectors, has been originally proposed
by Samuelson, Sukhorukov, and B\"uttiker (SSB). In Ref. \cite{BSS}
it was shown that while the current (and the noise) at each
particular detector is {\em insensitive} to the AB flux through
the interferometer, the current cross-correlations between the two
detectors do show AB oscillations. The latter are a direct
consequence of particle indistinguishability: Measuring current
signals at points c and d (cf. Fig.~1a) may be due to an electron
from a (b) absorbed at c (d), or at d (c). The product of the
amplitudes of these two processes is flux sensitive. Had the two
particles been distinguishable, it would have been possible to
conclude, for example, that the electron detected at c (d)
originated from a (b); only one of the above amplitudes survives,
hence flux insensitivity.

Trying to design an actual measurement of a two-electron
interferometry, it is important to understand which manipulations
can render the electrons distinguishable, in practice: suppress
the AB signal \cite{foot1}. Below we focus on two
interference-suppressing scenarios: (i) opposite spin
polarizations of the two sources, and (ii) entanglement of
electrons in one of the sources. We calculate how the visibility
in these two scenarios is reduced. In particular, our analysis
reveals that entangled electrons {\em vs} non-entangled electrons
act as (partially) distinguishable particles.

\begin{figure}[h]
\includegraphics[width=1.3\columnwidth]{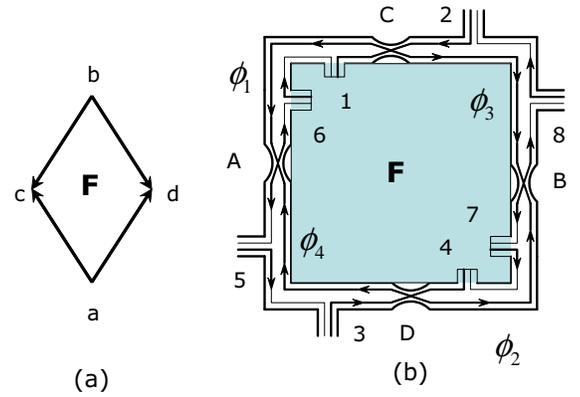}
\caption{{\protect\small A two-electron MZI. (a) Two electron
    sources (a,b)
    and two detectors (c,d), connected by chiral edges. An AB flux is
    threading the interferometer. The sensitivity of the current
    cross-correlation in the flux is the result of the indistinguishability of
    electrons emitted from a and b. Schematically, $\langle I_c I_d \rangle
    \propto \left\vert A_{b \to d} A_{a \to c} + A_{b \to c} A_{a \to d}
    \right\vert^2$, where $A_{b \to d}$ is the amplitude for transmission for
    $b$ to $d$. For example, the combination $A_{b\to d} A_{a \to c} A^*_{b
    \to c} A^*_{a \to d} = A_{b \to d} A_{d \to a} A_{a \to c} A_{c \to b}$ is
    phase sensitive. (b) A realistic setup with edge states~\cite{BSS},
    discussed in the text.}}
\label{fig:alpha}
\end{figure}

We first derive general expressions for the current and noise of a
generic multi-terminal system, in which electrons in the same or
different leads can be entangled. Let us consider a multi-terminal
system, with the leads emitting (possibly) entangled electrons. A
two-electron state in such a system has the form
\begin{equation} \label{entangled_state}
\hat F^{\dagger}_{\alpha\beta} (E, E') = \sum_{\sigma, \sigma'}
g^{*\alpha\beta}_{\sigma\sigma'} \hat a^{\dagger}_{\alpha\sigma} (E)
\hat a^{\dagger}_{\beta\sigma'} (E') \ ,
\end{equation}
where the indices $\alpha$ and $\beta$ label the leads, $\sigma$
and $\sigma'$ label the spin projections, and the coefficient $g$
is essentially the density matrix for the entangled electron
state. If the electrons are not entangled (full product state),
all coefficients $g$ equal one. For simplicity, we assume that each
lead supports a single channel: The energies $E$ and $E'$ describe
the longitudinal motion. Generalization to the multi-channel case is
straightforward.

The two-particle field operator depending on the coordinates,
times, and spin projections of both particles, is
\begin{eqnarray} \label{field}
& & \hat \Psi_{\sigma_1\sigma_2} (x_1 t_1; x_2 t_2) =
\frac{1}{2\pi\hbar v} \sum_{\beta\gamma} \int dE_1 dE_2 \nonumber \\
& & \times \phi_{\beta} (x_1, E_1) \phi_{\gamma} (x_2, E_2)
\exp(-iE_1t_1/\hbar - iE_2 t_2/\hbar) \nonumber \\
& & \times g^{\gamma\beta}_{\sigma_2\sigma_1} \hat a_{\beta\sigma_1}
\hat
a_{\gamma\sigma_2} \ ,
\end{eqnarray}
where for simplicity we take the electron velocities in all leads
equal to $v$, and $\phi_{\nu} (x)$ is the scattering state
originating from the lead $\nu$. The asymptotic form
for this function for $x \in \alpha$ is
\begin{displaymath}
\phi_{\nu} (x,E) = \delta_{\alpha\nu}e^{-ik(E)x} + s_{\alpha\nu} (E)
e^{ik(E)x} \ ,
\end{displaymath}
where, as common in the scattering approach, the coordinate $x$ is
counted from the center of the structure along each lead, and
$\hat s$ is the scattering matrix (below it will gain some more
indices). The scattering states are orthogonal, in discrete
notations $\int dx \phi^*_{\mu} (x, E) \phi_{\nu} (x, E') =
L\delta_{\mu\nu} \delta_{EE'}$, with $L$ being the size of the
system (it drops out of the final expressions) \cite{foot2}.

Now we produce the one-particle current operator in the lead
$\alpha$. It consists of the current operators produced by the
``first'' ($\hat I_1$) and the ``second'' ($\hat I_2$) particle. We
write
\begin{eqnarray*}
& & \hat I_{\alpha 1} (x_1, t_1) = -\frac{ie\hbar}{2m} \int dx_2
\sum_{\sigma_1 \sigma_2} \left\langle
\hat \Psi^{\dagger}_{\sigma_1\sigma_2} (x_1, t_1; x_2, t_2) \right. \\
& & \times \left. \frac{\partial}{\partial x_1} \hat \Psi_{\sigma_1
\sigma_2} (x_1, t_1; x_2, t_2) \right\rangle_2 + h.c\ ,  \ \ \ (WRONG)
\end{eqnarray*}
where $\langle \dots \rangle_2$ means quantum-mechanical averaging
over the state of the second particle. Using~\cite{foot3}
\begin{displaymath}
\left \langle \hat a^{\dagger}_{\lambda\sigma} (E) \hat a_{\nu\sigma'} (E')
\right\rangle = \delta_{\lambda\nu} \delta_{\sigma\sigma'} \delta (E - E')
f (E - \mu_{\lambda}) \ ,
\end{displaymath}
where $f$ is the equilibrium distribution function, and $\mu_{\lambda}$
is the corresponding chemical potential, we obtain for the
single-particle current operator
\begin{eqnarray*}
& & \hat I_{\alpha 1} (x, t) = -\frac{e}{2\pi\hbar} \frac{L}{2\pi\hbar
v} \int dE_1 dE_2
e^{i(E_1 - E_2)t/\hbar} \\
& & \times \sum_{\beta\gamma\delta}
\sum_{\sigma_1\sigma_2} A^{\alpha}_{\beta\delta} (E_1, E_2)
N_{\gamma\sigma_2}  g^{*\beta\gamma}_{\sigma_1\sigma_2}
g^{\delta\gamma}_{\sigma_1\sigma_2} \\
& & \times \hat a^{\dagger}_{\beta\sigma_1} (E_1) \hat
a_{\delta\sigma_1} (E_2) N_{\gamma\sigma_2} \ , \ \ (STILL \
WRONG)
\end{eqnarray*}
where we have introduced the following combination of the scattering
matrices \cite{Buttiker},
\begin{eqnarray} \label{scatmatrButtiker}
& & A^{\alpha}_{\beta\delta} (E_1, E_2) \equiv
\delta_{\alpha\beta}\delta_{\alpha\delta} e^{i[k(E_1)-k(E_2)]x} \\
& & - s^*_{\alpha\beta} (E_1) s_{\alpha\delta} (E_2)
e^{-i[k(E_1)-k(E_2)]x} \ ,
\end{eqnarray}
and the quantity $N_{\alpha\sigma} \equiv \int dE f_{\alpha\sigma} (E) dE$,
which is proportional to the ``number of particles'' with the spin
$\sigma$ emitted by the lead $\gamma$.

The expression for $\hat I_{\alpha 1}$ is obviously wrong: It
yields the
average current proportional to the number of {\em pairs of
particles}, not to the difference of the number of particles emitted
by each lead.  This is because our expression for the current operator
is an overcounting: Each pair of the particles is counted twice. Thus,
we need to add a normalization constant in the current
operator. This normalization constant is the number of pairs which a
particle incident from the lead $\gamma$ with the spin $\sigma_2$
forms, divided by the number of such particles. Note that this
constant must depend on $\gamma$ and $\sigma_2$. Let us now calculate
it,
\begin{eqnarray*}
& & \mbox{\# of pairs} = \int dx_1 dx_2 dE_1 dE_2 dE_3 dE_4
\sum_{\beta\sigma_1} \frac{1}{(2\pi\hbar v)^2} \\
& & \times \phi^*_{\beta} (x_1,
t_1, E_1) \phi^*_{\gamma} (x_2, t_2, E_2) \phi_{\gamma} (x_2, t_2,
E_3) \phi_{\beta} (x_1, t_1, E_4) \\
& & \times \left\vert g^{\beta\gamma}_{\sigma_1\sigma_2}
\right\vert^2 \left\langle a^{\dagger}_{\beta\sigma_1} (E_1)
a^{\dagger}_{\gamma\sigma_2} (E_2)
a_{\gamma\sigma_2} (E_3) a_{\beta\sigma_1} (E_4) \right\rangle \\
& & = \left( \frac{L}{2\pi\hbar v} \right)^2 \sum_{\beta\sigma_1}
\left\vert g^{\beta\gamma}_{\sigma_1\sigma_2} \right\vert^2
N_{\beta\sigma_1} N_{\gamma\sigma_2}
\end{eqnarray*}
and $\mbox{\# of particles} = (L/2\pi\hbar v) N_{\gamma\sigma_2}$.
Thus, including the normalization constant, we find for the current
operator of the first particle,
\begin{eqnarray} \label{curop}
& & \hat I_{\alpha 1} (x, t) = -\frac{e}{2\pi\hbar} \int dE_1 dE_2
e^{i(E_1 - E_2)t/\hbar} \nonumber \\
& & \times \sum_{\beta\gamma\delta} \sum_{\sigma_1\sigma_2}
\frac{\displaystyle{g^{*\beta\gamma}_{\sigma_1\sigma_2}
g^{\delta\gamma}_{\sigma_1\sigma_2}
N_{\gamma\sigma_2}}}{\displaystyle{\sum_{\nu\sigma} \left\vert
g^{\nu\gamma}_{\sigma\sigma_2} \right\vert^2 N_{\nu\sigma}}}
\nonumber \\
& & \times A^{\alpha}_{\beta\delta} (E_1, E_2) \hat
a^{\dagger}_{\beta\sigma_1}
(E_1) \hat a_{\delta\sigma_1} (E_2) N_{\gamma\sigma_2}
\ ,
\end{eqnarray}
and an identical expression for the current operator of the second
particle.

Next we use Eq. (\ref{curop}) to derive the average current,
\begin{eqnarray} \label{averagecurrent}
& & \left\langle I_{\alpha} \right\rangle =\left\langle I_{\alpha 1}
\right\rangle + \left\langle I_{\alpha 2} \right\rangle =
-\frac{e}{2\pi\hbar} \sum_{\gamma\sigma_2} \nonumber \\
& & \times
\frac{\displaystyle{\sum_{\beta\sigma_1} A^{\alpha}_{\beta\beta}
(\gamma\sigma_2)
\left\vert g^{\beta\gamma}_{\sigma_1\sigma_2} \right\vert^2
N_{\beta\sigma_1}
N_{\gamma\sigma_2}}}{\displaystyle{\sum_{\beta\sigma_1}\left\vert
g^{\beta\gamma}_{\sigma_1\sigma_2} \right\vert^2
N_{\beta\sigma_1}}} \ ,
\end{eqnarray}
where we assumed that the scattering matrices are energy independent,
and we added to them two additional indices, as is explained below.

Eq. (\ref{curop}) also yields current noise. To produce the expression
for the zero-frequency noise, we take into account that the currents of
the first and second particle are uncorrelated, and
averaging the product of four creation and annihilation operators
\cite{Buttiker},
\begin{eqnarray*}
& & \left\langle \hat a^{\dagger}_{\beta\sigma} (E_1)
\hat a_{\beta'\sigma'} (E_1') \hat a^{\dagger}_{\delta\sigma} (E_2)
\hat a_{\delta' \sigma'} (E_2') \right\rangle \\
& & - \left\langle
\hat a^{\dagger}_{\beta\sigma} (E_1) \hat a_{\beta'\sigma'} (E_1')
\right\rangle \left\langle \hat a^{\dagger}_{\delta\sigma} (E_2)
\hat a_{\delta' \sigma'} (E_2') \right\rangle \\
& & = \delta_{\beta\delta'}
\delta_{\beta'\delta} \delta_{\sigma\sigma'} \delta(E_1 - E_2')
\delta(E_1' - E_2) \\
& & \times f_{\beta\sigma} (E_1) \left( 1 - f_{\beta'\sigma}
(E_2) \right) \ ,
\end{eqnarray*}
we obtain
\begin{eqnarray} \label{currentnoise}
& & S_{\alpha\alpha'} = \frac{e^2}{2\pi\hbar} \sum_{\beta, \beta',
\gamma,
\gamma'} \sum_{\sigma_1, \sigma_2, \sigma_2'}
\frac{N_{\gamma\sigma_2}}{\displaystyle{\sum_{\beta\sigma_1}
\left\vert g^{\beta\gamma}_{\sigma_1\sigma_2} \right\vert^2
N_{\beta\sigma_1}}} \nonumber \\
& & \times
\frac{N_{\gamma'\sigma_2'}}{\displaystyle{\sum_{\beta\sigma_1}
\left\vert g^{\beta\gamma'}_{\sigma_1\sigma_2'} \right\vert^2
N_{\beta\sigma_1}}} \mbox{Re} \ \left[
g^{*\beta\gamma}_{\sigma_1\sigma_2} g^{\beta'
\gamma}_{\sigma_1\sigma_2} g^{*\beta'\gamma'}_{\sigma_1\sigma_2'}
g^{\beta\gamma'}_{\sigma_1\sigma_2'} \right] \nonumber \\
& & \times  A^{\alpha}_{\beta\beta'}
(\gamma, \sigma_2) A^{\alpha'}_{\beta'\beta} (\gamma', \sigma_2') \int
dE f_{\beta\sigma_1} (E) \left[ 1 - f_{\beta'\sigma_1} (E) \right]
\nonumber \\
& & +
\left( \alpha \leftrightarrow \alpha' \right) \ .
\end{eqnarray}

Eqs. (\ref{averagecurrent}) and (\ref{currentnoise}) generalize
standard expressions for the multi-terminal current and noise
to the case of entangled electrons. It is easy to
check that if electrons are not entangled and are in the product
state, the coefficients $g$ are equal one for all values of the
arguments, and these expressions reduce to the standard one-particle
formulas \cite{Buttiker}.

The problem with Eqs. (\ref{averagecurrent}) and
(\ref{currentnoise}) is that currents are not automatically
conserved, $\sum_{\alpha} I_{\alpha} \ne 0$. Also, if all leads
have equal chemical potentials, non-vanishing currents can be
still generated according to Eq. (\ref{averagecurrent}). Usually,
requirements of current conservation and gauge invariance
(currents are unchanged if the chemical potentials of all
reservoirs are shifted simultaneously) are guaranteed by the
unitarity of the scattering matrix. In our case, they are
satisfied {\em automatically} provided we choose the scattering
matrix to obey
\begin{eqnarray} \label{unitarity}
& & \sum_{\mu} s^*_{\mu\nu} (\gamma, \sigma_2) s_{\mu\nu'} (\gamma',
\sigma_2') = \delta_{\nu\nu'} \ ;\\
& & \sum_{\nu}
g^{*\nu\gamma}_{\sigma\sigma_2}g^{\nu\gamma'}_{\sigma\sigma_2'}
s^*_{\mu\nu} (\gamma, \sigma_2) s_{\mu'\nu} (\gamma', \sigma_2') =
\delta_{\mu\mu'} g^{*\mu\gamma}_{\sigma\sigma_2}
g^{\mu\gamma'}_{\sigma\sigma_2'}  \nonumber \ .
\end{eqnarray}

Next, we specialize to the eight-terminal setup with edge states
suggested by SSB \cite{BSS}. We will
always bias sources $2$ and $3$ with the same voltage $V$, and measure
the average current through lead $5$, and the cross-correlation
between currents at $5$ and $8$. We are interested in the linear
in $V$ effects. Assuming that $eV$ is much smaller that the Fermi energy in
the leads, we can in the leading order take all quantities $N_{\beta\sigma}$
to be the same (voltage independent). One has then
\begin{equation} \label{curSSB}
\left\langle I_5 \right\rangle = -\frac{e^2 V}{2 \pi \hbar}
\sum_{\gamma\sigma_2} \frac{1}{\displaystyle{\sum_{\beta\sigma_1}
\left\vert g^{\beta\gamma}_{\sigma_1\sigma_2} \right\vert^2 }}
\displaystyle{\sum_{\beta = 2,3}\sum_{\sigma_1} \left\vert s_{5\beta}
(\gamma, \sigma_2) \right\vert^2}
\end{equation}
and
\begin{eqnarray} \label{noiseSSB}
& & S_{58} = -\frac{e^3 \vert V \vert}{\pi\hbar} \sum_{\gamma\gamma'}
\sum_{\sigma_1 \sigma_2 \sigma_2'}
\frac{1}{\displaystyle{\sum_{\beta\sigma_1}
\left\vert g^{\beta\gamma}_{\sigma_1\sigma_2} \right\vert^2 }}
\frac{1}{\displaystyle{\sum_{\beta\sigma_1}
\left\vert g^{\beta\gamma'}_{\sigma_1\sigma_2'} \right\vert^2 }}
\nonumber \\
& & \times \displaystyle{ \left\vert \sum_{\beta=2,3}
g^{*\beta\gamma}_{\sigma_1\sigma_2}
g^{\beta\gamma'}_{\sigma_1\sigma_2'} s^*_{5\beta} (\gamma, \sigma_2)
s_{8\beta} (\gamma', \sigma_2') \right\vert^2 } \ ,
\end{eqnarray}
where we have used the ``unitarity'' conditions (\ref{unitarity}).

Now we apply these expressions to the different states emitted by the
reservoirs. The idea is to investigate under which conditions
electrons emitted from sources 2 and 3 are ``painted'', at least
partially, ``red'' and ``blue'' respectively, hence suppression of the
flux sensitive (coherent) term of the correlation.

(i) {\em Full product state}, $g^{\beta\gamma}_{\sigma\sigma'} =
1$. This is the case considered by SSB. Choosing $s_{52} =
t_At_Ce^{i\phi_1}$, $s_{53} = r_Ar_De^{i\phi_4}$,
$s_{82}=r_Br_Ce^{i\phi_3}$, and $s_{83} = t_Bt_D e^{i\phi_2}$,
where $t$ and $r$ are the transmission and reflection amplitudes
of the corresponding QPC's (Fig.~1), we obtain
\begin{equation} \label{avcurfull}
\left\langle I_5 \right\rangle = -\frac{e^2V}{2\pi\hbar} \left(
\left\vert s_{52} \right\vert^2 + \left\vert s_{53} \right\vert^2
\right) = -\frac{e^2V}{2\pi\hbar} \left( T_AT_C + R_AR_D \right) \ ,
\end{equation}
where $T$ and $R$ are corresponding transmission and reflection
probabilities, and
\begin{eqnarray} \label{noisefull}
S_{58} & = & -\frac{2e^3 \vert V \vert}{\pi\hbar} \left\vert
s^*_{52}s_{82} + s^*_{53}s_{83} \right\vert^2 \\
& = & -\frac{2e^3 \vert V
\vert}{\pi\hbar} \left\vert \sqrt{T_AT_CR_BR_C} e^{i\theta} +
\sqrt{T_BT_DR_AR_D} \right\vert^2 \nonumber \ ,
\end{eqnarray}
where $\theta$ is a linear combination of the phases of transmission
and reflection amplitudes and the Aharonov-Bohm phase. In accordance with
SSB, the average current is not sensitive to AB effect, since not a
single electron trajectory encircles the magnetic flux. By contrast,
noise is a two-particle phenomenon, and the setup is designed in such
a way that the cross-correlations are sensitive to the AB flux. The
visibility of the noise oscillations is
\begin{equation} \label{visibility}
v = \frac{2\sqrt{T_AT_BT_CT_DR_AR_BR_CR_D}}{T_AT_CR_BR_C +
T_BT_DR_AR_D}
\end{equation}
and becomes unity (ideal oscillations) provided the setup is symmetric:
all transmission and reflection probabilities equal $1/2$. This is the
result of SSB.

(ii) {\em One source is spin polarized}. Imagine that states are
still not entangled, but source $2$ is spin-polarized: It only
emits and absorbs spin-up electrons. The coefficients $g$ are all
equal to one, with the exception of
$g^{2\gamma}_{\downarrow\sigma_2} = g^{\beta
2}_{\sigma_1\downarrow} = 0$. We need now to choose the scattering
matrices in such a way so that they obey the unitarity condition
(\ref{unitarity}). This is easily done intuitively. For spin-up
electrons, the constraints are the same as for the full product
state, and thus the scattering for spin-up electrons is described
by the matrix $s$. The scattering matrix for spin-down electrons,
which we denote $\tilde s$, is constrained by the condition
\begin{displaymath}
\sum_{\nu \ne 2} \tilde s^*_{\mu\nu} \tilde s_{\mu'\nu} =
\delta_{\mu\mu'} \ , \tilde s_{\mu 2} = 0 \ .
\end{displaymath}

We obtain the average current,
\begin{equation} \label{curspinpolar}
\left\langle I_5 \right\rangle = -\frac{e^2V}{2\pi\hbar} \left(
\left\vert s_{52} \right\vert^2 + \left\vert s_{53} \right\vert^2 +
\left\vert \tilde s_{53} \right\vert^2 \right) \ ,
\end{equation}
and current noise,
\begin{equation} \label{noisespinpolar}
S_{58} = -\frac{e^3 \vert V \vert}{\pi\hbar} \left( \left\vert
s^*_{52}s_{82} + s^*_{53}s_{83} \right\vert^2 + \left\vert
\tilde s^*_{53} \tilde s_{83} \right\vert^2 \right) \ .
\end{equation}
The first term in the brackets, similarly to Eq.
(\ref{noisefull}), contains both phase-insensitive and AB terms.
The second term is insensitive to the AB phase. Thus, the total
visibility in the noise is reduced by the presence of this second
term. The visibility depends on the choice of the scattering
matrices $\tilde s$, in particular,  if the setup is symmetric,
and $\tilde s_{53} = 0$ (unrealistic case), the oscillations are
still ideal ($v = 1$). This is because in this case a spin-down
electron, originating from $3$, has to go to $8$ with certainty,
and thus does not change the current noise. On the other hand, the
natural choice would be $\vert s_{52} \vert = \vert s_{53} \vert =
\vert \tilde s_{52} \vert$. In this case, which we call {\em
optimal}, the visibility equals $v = 2/3$.

In the same way, we can treat a setup where both sources, $2$ and $3$, are
polarized. If they
are polarized in the same direction (spin-up), the visibility is
the same as for full product states (both current and noise are reduced by the
factor of $2$ since now only spin-up states contribute). Provided one source
is spin-up polarized and the other one is spin-down, the visibility vanishes:
There are no AB oscillations in this case, since one can say with certainty
from what source each electron has originated.

(iii) {\em Entangled electrons from one source}. Pairs in {\em
triplet} $s_z = 0$ state are emitted from the lead $2$,
\begin{displaymath}
\vert 2\uparrow \rangle \vert 2\downarrow \rangle + \vert 2\downarrow
\rangle \vert 2\uparrow \rangle  \ ,
\end{displaymath}
all other leads are in the full product state. This means that all the
coefficients $g$ are equal to one except for
$g^{22}_{\uparrow\uparrow} = g^{22}_{\downarrow\downarrow} = 0$. Now
we are obliged to choose scattering matrices which depend on the
second electron in the pair --- the ``spouse'', otherwise the
unitarity (\ref{unitarity}) conditions can not be fulfilled. We
choose
\begin{eqnarray*}
s_{\alpha\nu, \sigma} (\gamma, \sigma_2) = \left\{ \begin{array}{lr}
s_{\alpha\nu} \ , & \gamma \ne 2 \ \mbox{or} \ \gamma=2, \sigma \ne
\sigma_2
\\
\tilde s_{\alpha\nu} \ , & \gamma = 2, \sigma = \sigma_2 \end{array}
\right. \ .
\end{eqnarray*}
Note that this choice (affecting the visibility) is arbitrary and definitely
not unique. Our matrices now, in addition to what we have already seen, obey
\begin{displaymath}
\sum_{\nu \ne 2} \tilde s^*_{\mu\nu} s_{\mu'\nu} = \sum_{\nu \ne 2}
s^*_{\mu\nu} \tilde s_{\mu'\nu} =
\delta_{\mu\mu'} \ .
\end{displaymath}
We can again calculate the average current,
\begin{equation} \label{curtriplet}
\left\langle I_5 \right\rangle = -\frac{e^2V}{2\pi\hbar} \left\{ \left(
\frac{7}{8} + \frac{1}{15} \right) \left(
\left\vert s_{52} \right\vert^2 + \left\vert s_{53} \right\vert^2
\right) + \frac{1}{15}\left\vert \tilde s_{53} \right\vert^2 \right\}
\ ,
\end{equation}
and the current noise,
\begin{eqnarray} \label{noisetriplet}
& & S_{58} = -\frac{2e^3 \vert V \vert}{\pi\hbar} \left\{ \left(
\frac{14^2}{16^2} + 2\frac{14}{15\cdot 16} + \frac{1}{15^2} \right)
\right. \\
& & \left. \times \left\vert s^*_{52}s_{82} + s^*_{53}s_{83}
\right\vert^2 + \left(
\frac{14}{15 \cdot 16} + \frac{1}{15^2} \right) \right. \nonumber \\
& & \left. \times \left( \left\vert
\tilde s^*_{53} s_{83} \right\vert^2 + \left\vert
s^*_{53} \tilde s_{83} \right\vert^2 \right) +
\frac{1}{15^2} \left\vert
\tilde s^*_{53} \tilde s_{83} \right\vert^2 \right\} \nonumber \ .
\end{eqnarray}
Again, phase-dependent contributions are only found in the first
term in the braces. All other terms are phase insensitive and thus
reduce the visibility. In particular, with the same optimal choice
of scattering matrices, $\vert s_{52} \vert = \vert s_{53} \vert =
\vert \tilde s_{52} \vert$, the visibility is reduced to $0.93$.

(iv) Pairs in {\em singlet} state are emitted from the lead $2$,
\begin{displaymath}
\vert 2\uparrow \rangle \vert 2\downarrow \rangle - \vert 2\downarrow
\rangle \vert 2\uparrow \rangle  \ ,
\end{displaymath}
all other leads are in the full product state. This means that all
the coefficients $g$ are equal one except for
$g^{22}_{\uparrow\uparrow} = g^{22}_{\downarrow\downarrow} = 0$,
and $g^{22}_{\downarrow\uparrow}=-1$. We choose the scattering
matrices in the following way (this choice is again arbitrary),
\begin{eqnarray} \label{scatsinglet}
s_{\alpha\nu, \sigma} (\gamma, \sigma_2) = \left\{ \begin{array}{lr}
s_{\alpha\nu} \ , & \gamma \ne 2 \mbox{or} \ \gamma=2, (\sigma\sigma_2) =
(\uparrow\downarrow)
\\
\tilde s_{\alpha\nu} \ , & \gamma = 2, \sigma = \sigma_2 \\
\bar s_{\alpha\nu} \ , & \gamma = 2, (\sigma\sigma_2) =
(\downarrow\uparrow)  \end{array}
\right. \ .
\end{eqnarray}
The matrix $\bar s$ must then obey the following constraints,
\begin{eqnarray*}
& & \sum_{\nu} \bar s^*_{\mu\nu} \bar s_{\mu'\nu} = \sum_{\nu}
s^*_{\mu\nu} \bar s_{\mu'\nu} = \sum_{\nu}
\bar s^*_{\mu\nu} s_{\mu'\nu} \\
& & = \sum_{\nu \ne 2} \bar s^*_{\mu\nu}
\tilde s_{\mu'\nu} = \sum_{\nu \ne 2} \tilde s^*_{\mu\nu}
\bar s_{\mu'\nu} =
\delta_{\mu\mu'} \ .
\end{eqnarray*}
With this choice of scattering matrices, we obtain the same current
and current noise (and consequently the same visibility) as for the
triplet $s_z = 0$ state. We should recall however that this
conclusion depends on the choice of scattering matrices
(\ref{scatsinglet}): for instance, on the fact, that $s_{\alpha\nu,
\sigma} (2, \sigma)$ is the same for single and triplet entangled
states.

\begin{table}
\begin{tabular}{lc}
Scenario & Optimal visibility \\
$2$, $3$: product states & $1$ \\
$2$: product, not polarized; $3$: polarized $\uparrow$ & $2/3$ \\
$2$, $3$: polarized $\uparrow$ & $1$ \\
$2$: polarized $\uparrow$, $3$: polarized $\downarrow$ & $0$ \\
$2$: entangled, singlet; $3$: product & $0.93$ \\
$2$: entanglet, triplet; $3$: product & $0.93$
\end{tabular}
\caption{Reduction of the visibility for different scenarios and otherwise
  optimal conditions.}
\end{table}

In summary, we have defined and analyzed various scenarios for which
single-particle amplitudes maintain their coherence (as would be
manifest in a MZ interferometry measurement), yet (partial)
distinguishability of electrons emitted from different sources
suppresses the flux sensitivity of the two-particle cross-correlation
function (Table 1). Our analysis demonstrates how two-particle entanglement can
give rise to a behavior akin to a ``dephasing'' of a two-particle
interferometry.

We acknowledge useful discussions with M.~B\"uttiker, M.~Heiblum, and
I.~Neder. This work was supported by the ISF of the Israel Academy of
Sciences, by the US -- Israel BSF, by the Minerva Einstein Center (BMBF), and
by the Transnational Access Program RITA-CT-2003-506095 at the Weizmann
Institute of Science.


\begin{thebibliography}
\bibitem{foot2} It will be more natural and
convenient to use continous notations for the energy, $\int dx
\phi^*_{\mu} (x, E) \phi_{\nu} (x, E') = 2\pi\hbar v\delta_{\mu\nu}
\delta (E-E')$. In particular, one can also write the ``square'' of the
delta-function in continuous notations,
$\delta^2 (E - E') = (L/2\pi v) \delta (E - E')$.
\end{thebibliography}
\end{document}